\documentclass{amsart}

\usepackage{amsbsy,amssymb,amscd,amsfonts,latexsym,amstext,delarray,
amsmath,graphicx}
\input xypic
\usepackage{color}

\numberwithin{equation}{section}

\def\C{{\mathbb C}}

\renewcommand{\H}{{\mathbb H}}

\def\Z{{\mathbb Z}}
\def\R{{\mathbb R}}

\def\cA{{\mathcal A}}
\def\cB{{\mathcal B}}
\def\cC{{\mathcal C}}

\def\cF{{\mathcal F}}
\def\cG{{\mathcal G}}
\def\cH{{\mathcal H}}

\def\cM{{\mathcal M}}

\def\cR{{\mathcal R}}
\def\cS{{\mathcal S}}

\def\cV{{\mathcal V}}

\def\Tr{{\rm Tr}}

\def\GL{{\rm GL}}

\def\fa{{\mathfrak{a}}}
\def\fb{{\mathfrak b}}
\def\fc{{\mathfrak c}}
\def\fd{{\mathfrak d}}
\def\fe{{\mathfrak e}}

\def\mass{Y}

\def\cancel#1#2{\ooalign{$\hfil#1\mkern1mu/\hfil$\crcr$#1#2$}}

\def\dirac{\mathpalette\cancel\partial}

\title[Asymptotic safety, hypergeometric functions and Higgs mass]
{Asymptotic safety, hypergeometric
functions, and the Higgs mass in spectral action models}
\author{Christopher Estrada and Matilde Marcolli}
\address{Division of Physics, Mathematics and Astronomy, 
Mail Code 253-37, Caltech, 1200 E.~California Blvd. Pasadena, CA 91125, USA}
\email{c.estrada@caltech.edu}
\email{matilde@caltech.edu}

\begin{document}
\maketitle

\begin{abstract}
We study the renormalization group flow for the Higgs self coupling in
the presence of gravitational correction terms. We show that the resulting
equation is equivalent to a singular linear ODE, which has explicit
solutions in terms of hypergeometric functions. We discuss the implications
of this model with gravitational corrections on the Higgs mass estimates
in particle physics models based on the spectral action functional.
\end{abstract}

\bigskip
\bigskip

{\em Ella est\'a en el horizonte. Me acerco dos pasos, ella se aleja dos pasos. 
Camino diez pasos y el horizonte se corre diez pasos m\'as all\'a. 
Por mucho que yo camine, nunca la alcanzar\'e. 
?`Para que sirve la utop\'ia? 
Para eso sirve: para caminar. } 

\smallskip

(Eduardo Galeano)

\section{Introduction}

A realistic Higgs mass estimate of 126 GeV was obtained 
by Shaposhnikov and Wetterich in \cite{ShaWett}, 
based on a renormalization group analysis, using the functional renormalization
group equations (FRGE) method of \cite{Wett} for gravity coupled to matter. 
In this setting, the renormalization group equations (RGE) for the matter sector 
acquire correction terms coming from the gravitational parameters, which
are expressible as additional terms in the beta functions which depend on
certain parameters: the anomalous dimensions $a_x$ and the scale
dependence $\rho_0$ of the Newton constant. These gravitational
correction terms to the RGE make the matter couplings asymptotically free,
giving rise to a ``Gaussian matter fixed point" (see \cite{NaPe}, \cite{PePe}).

In this paper we carry out a more detailed mathematical analysis of the
RGE for the Higgs self-coupling, in this asymptotic safety scenario with
anomalous dimensions, using the standard approximation that keeps
only the dominant term in the Yukawa coupling matrices coming from
the top quark Yukawa coupling. We show that the resulting equations 
have explicit solutions in closed form, which can be expressed in terms
of hypergeometric functions.

We discuss the implications of using this RGE flow on the particle
physics models based on the spectral action functional. In particular,
one can obtain in this way a realistic Higgs mass estimate without 
introducing  any additional field content to the model (see the recent
\cite{CCnew} for a different approach based on a coupling with a
scalar field), but the fact that the RGE with anomalous dimensions
lead to a ``Gaussian matter fixed point" at high energies requires a
reinterpretation of the geometric constraints at unification energy
imposed by the geometry of the spectral action models, as in \cite{CCM}.

The paper is organized as follows. In \S \ref{NCGsec} we review the
basic formalism of models of matter coupled to gravity based on
noncommutative geometry and the spectral action functional, and in
particular of the model described in \cite{CCM} and in Chapter 1 of \cite{CoMa}.
In \S \ref{RGE0sec} we review the use of renormalization group analysis
in these models, in the case where the renormalization group equations
used are those of the Minimal Standard Model (MSM) or of the extension
with right handed neutrinos with Majorana mass terms ($\nu$MSM).
We describe the usual approximations to the full system of equations,
in view of our later use of the same approximations in the presence of
gravitational corrections. We also show how it is natural to think, in these
models, of the RG flow as a flow of the finite noncommutative geometry,
or equivalently a flow on the moduli space of Dirac operators on the
finite geometry.  In \S \ref{RGEsec} we show that it makes sense to
apply Wetterich's functional renormalization group equations (FRGE)
developed in \cite{Wett} to the spectral action. Using the high
precision of the approximation of the spectral action by an action
functional for the Higgs field non-minimally coupled to gravity, this
leads to renormalization group equations of the type derived in
\cite{NaPe}, \cite{PePe} for matter coupled to gravity, within the
asymptotic safety scenario, with a Gaussian matter fixed point.
These renormalization group equations differ from the usual ones
described in \S \ref{RGE0sec} by the presence of gravitational
correction terms with anomalous dimensions. In \S \ref{RGEgravSec}
we study the resulting RGE with anomalous dimensions, under the
same approximations described in \S \ref{RGE0sec}. We show that
one can give explicit solutions for the resulting equations for the
top Yukawa coupling and the Higgs self-coupling and that the
latter is expressible in terms of the Gauss hypergeometric function ${}_2 F_1(a,b,c,z)$.
We show that one can fit the Shapshnikov--Wetterich Higgs mass estimate
within this setting. In \S \ref{UnifSec} we discuss the effect of these modified
RGE on the constraints imposed by the geometry of the model on the
boundary conditions at unification.

\section{The spectral action and the NCG models of particle physics}\label{NCGsec}

\smallskip
\subsection{General geometric framework of NCG models}
Noncommutative geometry models for particle physics coupled to gravity
are based on enriching the ordinary four-dimensional spacetime manifold $X$
to a product  $X \times F$ (more generally, a nontrivial fibration) with
``extra dimensions" $F$ consisting of a noncommutative space. The noncommutative
geometry $F$ is described by a {\em spectral triple}, that is, data of the form
$(\cA,\cH,D)$, where $\cA$ is an involutive algebra acting on a Hilbert space $\cH$,
and a Dirac operator $D$ on $\cH$ satisfying a compatibility condition with $\cA$,
that commutators $[D,a]$ are bounded operators, while $D$ itself is a densely
defined self-adjoint operator on $\cH$ with compact resolvent. The spectral triple
is additionally endowed with a $\Z/2\Z$-grading $\gamma$ on $\cH$, with
$[\gamma,a]=0$ and $D\gamma =-\gamma D$, and a
real structure $J$, which is an anti-linear isometry on $\cH$ satisfying
$J^2 = \varepsilon$, $JD = \varepsilon' DJ$, and $J\gamma = \varepsilon'' \gamma J$,
where $\epsilon, \epsilon', \epsilon''$ are signs $\pm 1$ that determine the KO-dimension
of the noncommutative space. The real structure also satisfies compatibility
conditions with the algebra representation and the Dirac operator, given by
$[a,b^0] = 0$ for all $a,b\in \cA$ with $b^0 = J b^* J^{-1}$ and the {\em order
one condition} for the Dirac operator: $[[D,a],b^0] = 0$ for all
$a,b\in \cA$.

In these models one generally assumes that the
noncommutative space $F$ is {\em finite}, which means that $\cA$ and $\cH$
are finite dimensional. In this case, the spectral triple data become just 
linear algebra data and the conditions listed above reduce to equations
$D^*=D$, $[a,b^0] = 0$, and $[[D,a],b^0] = 0$, to be solved with the constraints
$[\gamma,a]=0$, $D\gamma =-\gamma D$, $J^2 = \varepsilon$, 
$JD = \varepsilon' DJ$, and $J\gamma = \varepsilon'' \gamma J$.

\smallskip
\subsection{$\nu$MSM coupled to gravity from NCG}
In \cite{CCM} a particle physics model coupled to gravity is constructed
with this method. It is shown (see also \S 1 of \cite{CoMa}) that it recovers
an extension of the minimal Standard Model with right handed neutrinos and
Majorana mass terms. The {\em ansatz} algebra for the finite geometry $F$,
in the model of \cite{CCM} is taken to be
$\cA_{LR}= \C \oplus \H_L \oplus \H_R \oplus M_3(\C)$. 
The main step of the construction of \cite{CCM} are the following.
A representation of the algebra $\cA_{LR}$ is obtained by taking the sum $\cM_F$ of all the inequivalent irreducible $\cA_{LR}$-bimodules (with an {\em odd} condition that refers to
the action of a natural involution in the algebra, see \cite{CCM}). 
The number $N$ of particle generations is {\em not} predicted by the 
model and is assigned by taking as Hilbert space of the finite spectral 
triple $\cH_F =\oplus^N \cM_F$. 
This representation space provides the fermion fields content of the particle physics model.
The left-right chirality symmetry of $\cA_{LR}$ is broken spontaneously by 
the order one condition of the Dirac operator. This selects a maximal subalgebra 
on which the condition holds, while still allowing the Dirac operator to mix the 
matter and antimatter sectors. The subalgebra is of the form
$\cA_F=\C \oplus \H \oplus M_3(\C)$. The resulting noncommutative space $F$ 
is metrically zero dimensional but with KO-dimension $6$.
The real structure involution $J_F$ exchanges matter and antimatter, and the grading 
$\gamma_F$ distinguishes left and right chirality of particles.
There is a complete classification of all possible Dirac operators on 
this finite geometry and the parameters of the particle physics model that
include Yukawa parameters $Y$ (masses and mixing angles) and the Majorana 
mass terms $M$ of the right handed neutrinos geometrically arise as coordinates 
on the moduli space of Dirac operators on the finite geometry $F$.
The boson fields of the model arise as fluctuations of the Dirac operator $D\mapsto D_A$,
with the gauge bosons corresponding to fluctuation in the horizontal (manifold)
directions and the Higgs field arising as fluctuations in the vertical (noncommutative)
direction of the product space $X\times F$. In recent supersymmetric versions
of the NCG model \cite{BroSuij}, the supersymmetric partners of the Standard
Model fermions also arise as vertical fluctuations of the Dirac operator, 
for a different choice of the finite geometry $F$.

The field content of the model, including the neutrino sector, agrees with 
the $\nu$MSM model discussed in \cite{Shap1}, \cite{Shapo}, \cite{ShapoTka}. 
This is an extension of
the minimal Standard Model (MSM) by right handed neutrinos with
Majorana masses. In addition to the parameters of the MSM there are
additional real parameters that correspond to 
Majorana neutrino masses, and additional Yukawa coupling parameters for the
lepton sector given by Dirac neutrino masses, mixing angles, and 
CP-violating phases. The resulting model has a total number of 31 
real parameters coming from the moduli space of Dirac operators
on the finite geometry $F$ (which account for all the parameters
listed here) plus three coupling constants.
A significant difference with respect to the $\nu$MSM model
lies in the fact that this NCG model has a unification energy.

\smallskip
\subsection{The spectral action functional}
The spectral action functional for the model of \cite{CCM} is then given by
\begin{equation}\label{SAfermions}
 \Tr(f(D_A/\Lambda))  + \frac
12\,\langle\,J\,\tilde\xi,D_A\,\tilde\xi\rangle, 
\end{equation}
where the first term gives rise, in the asymptotic expansion, to all
the bosonic terms of the particle physics Lagrangian, coupled
(non-minimally) to gravity, and the second term gives the fermionic
terms, and their interactions with bosons. The variables $\tilde\xi$ are
fermions in the representation $\cH_F \otimes L^2(X,S)$, with $S$ the
spinor bundle on $X$, viewed as anticommuting Grassmann variables. 
The asymptotic expansion for large $\Lambda$
of the spectral action \cite{CC2} is of the form
\begin{equation}\label{SpActAsympt}
 \Tr(f(D/\Lambda))\sim \sum_{k\in {\rm DimSp^+}} f_{k} \Lambda^k {\int\!\!\!\!\!\!-} |D|^{-k} + f(0) \zeta_D(0)+ o(1),
\end{equation}
where $f_k= \int_0^\infty f(v) v^{k-1} dv$ and $f_0=f(0)$ are the momenta of the test function $f$, 
and where the integration ${\int\!\!\!\!\!\!-} a |D|^{-k}$ is defined by the residues of the family 
of zeta functions $\zeta_D (s) = \Tr (|D|^{-s})$ at
the positive points of the {\em dimension spectrum} of the spectral triple,
that is, their set of poles.

The asymptotic expansion \eqref{SpActAsympt}, as 
computed in \cite{CCM}, gives terms of the form
\begin{equation}\label{largeLambda}
\begin{array}{rl}
\Tr(f(D_A/\Lambda)) \sim
& \displaystyle{ \frac{1}{\pi^2}(48\,f_4\,\Lambda^4-f_2\,\Lambda^2\,\fc+\frac{
f_0}{4}\,\fd) }\,\int \,\sqrt g\,d^4 x \\[3mm]
    +& \, \displaystyle{
     \frac{96\,f_2\,\Lambda^2 -f_0\,\fc}{ 24\pi^2} }\, \int\,R
 \, \sqrt g \,d^4 x  \\[3mm]
    +& \, \displaystyle{
    \frac{f_0 }{ 10\,\pi^2} } \int\,(\frac{11}{6}\,R^* R^* -3 \, C_{\mu
\nu \rho \sigma} \, C^{\mu \nu \rho \sigma})\, \sqrt g \,d^4 x \\[3mm]
 +&  \, \displaystyle{
 \frac{(- 2\,\fa\,f_2
  \,\Lambda^2\,+ \,\fe\,f_0 )}{ \pi^2} } \int\,  |\varphi|^2\, \sqrt g \,d^4 x 
\\[3mm]
    +&  \, \displaystyle{
    \frac{f_0 \fa}{ 2\,\pi^2} } \int\,  |D_{\mu} \varphi|^2\, 
\sqrt g \,d^4 x \\[3mm]
   -&  \displaystyle{
   \frac{f_0 \fa}{ 12\,\pi^2} } \int\, R \, |\varphi|^2 \, \sqrt g \,d^4 x
 \\[3mm]
    +&\, \displaystyle{ \frac{f_0 \fb}{ 2\,\pi^2} } \int |\varphi|^4 \, \sqrt g \,d^4 x \\[3mm]
+& \, \displaystyle{ \frac{f_0 }{ 2\,\pi^2} } \int\,(g_{3}^2 \, G_{\mu \nu}^i \, 
G^{\mu \nu i} +  g_{2}^2 \, F_{\mu
\nu}^{ \alpha} \, F^{\mu \nu  \alpha}+\, \frac{5}{ 3} \,
g_{1}^2 \,  B_{\mu \nu} \, B^{\mu \nu})\, \sqrt g \,d^4 x ,
\end{array}
\end{equation}
where $R^*R^*$ is a topological non-dynamical term that integrates to a multiple of the
Euler characteristic $\chi(X)$, and $C_{\mu \nu \rho \sigma}$ is the Weyl curvature
tensor of conformal gravity. The Higgs field is non-minimally coupled to gravity through
the term $R |\varphi |^2$. The first two terms provide cosmological and Einstein--Hilbert
gravitational terms and the remaining terms are the kinetic and self-interaction terms
of the Higgs and the Yang--Mills terms of the gauge bosons. The coefficients of these
various terms depend on the parameters of the model and fix boundary conditions
and relations between parameters at unification energy. The constants $f_0$, $f_2$, $f_4$
(momenta of the test function) are free parameters of the model: $f_0$ is related 
to the value of the coupling constants at unification, while the remaining two parameters 
$f_2$ and $f_4$ enter in the expressions for the effective gravitational and cosmological
constants of the model. The parameters $\fa,\fb,\fc,\fd,\fe$ are functions of the 
Yukawa parameters and Majorana mass terms of the model:
\begin{equation}\label{abcde}
 \begin{array}{rl}
  \fa =& \,\Tr(\mass_{\nu}^\dag \mass_{\nu}+\mass_{e}^\dag \mass_{e}
  +3(\mass_{u}^\dag \mass_{u}+\mass_{d}^\dag \mass_{d})) \\[2mm]
  \fb =& \,\Tr((\mass_{\nu}^\dag \mass_{\nu})^2+(\mass_{e}^\dag \mass_{e})^2+3(\mass_{u}^\dag \mass_{u})^2+3(\mass_{d}^\dag \mass_{d})^2) \\[2mm]
  \fc =& \Tr(M M^\dag)  \\[2mm]
  \fd =& \Tr((M M^\dag)^2)  \\[2mm]
  \fe =& \Tr(M M^\dag \mass_{\nu}^\dag \mass_{\nu}) .
\end{array}
\end{equation}

\section{Renormalization group equations in NCG models}\label{RGE0sec}

It is customary, within the various NCG models of particle physics
coupled to gravity, to derive low energy estimates by assigning
boundary conditions at unification compatible with the geometric
constraints and running the renormalization group flow towards
lower energies. The beta functions that give the RGE
equations are imported from the relevant particle physics model.
Thus, for exampls, the estimates derived in   \cite{CaIoSchu} (for the
Connes--Lott model), or in \cite{CaIoKaSchu} (for NC Yang--Mills),
or in \cite{Oku} and, more recently, in \cite{CCM} are all based on the 
use of the RGE for the MSM. In the case of \cite{CCM} an estimated 
correction term to the running for the top quark is introduced coming from 
the Yukawa coupling term for the $\tau$ neutrino, which, in the
presence of the Majorana mass terms, becomes of comparable magnitude.
In \cite{MaKol} and \cite{MaPie}, the same NCG model of \cite{CCM}
is analyzed using the full RGE analysis for the extension of the
MSM by right handed neutrinos with Majorana masses, using
the technique derived in \cite{AKLRS},
giving rise to different effective field theories in between the
different see-saw scales.

\medskip
\subsection{RGE equations for the $\nu$MSM}

The particle physics content of the CCM model \cite{CCM} is an extension of
the minimal standard model (MSM) by right handed neutrinos with Majorana
mass terms, which, as we recalled above, is a $\nu$MSM model with
unification energy. Renormalization
group equations (at one loop) for this type of extension of the minimal standard model
are described in \cite{AKLRS}, while the renormalization group equations of
the MSM are known at one and two loops \cite{ACKPRW}.

The renormalization group equations for the $\nu$MSM are a system of
ordinary differential equations in unknown functions $x=(x_i)$ of the form
\begin{equation}\label{RGEx}
\partial_t x_i(t) = \beta_{x_i}(x(t)),
\end{equation}
in the variable $t =\log(\Lambda/M_Z)$,
where the beta functions for the various parameters (coupling constants,
Yukawa parameters, and Higgs quatric coupling) are given by
\begin{equation}\label{betaSMg}
\beta_1= \frac{41}{96 \pi^2}\, g_1^3, \ \ \ \  
\beta_2= - \frac{19}{96 \pi^2}\, g_2^3, \ \ \ \
\beta_3= -\frac{7}{16 \pi^2} g_3^3, 
\end{equation}
for the three coupling constants, while the beta functions for the
Yukawa parameters are of the form
\begin{equation}\label{RGEYu}
 16 \pi^2 \, \,  \beta_{\mass_u} =  
\mass_u(\frac{3}{2} \mass_u^\dag \mass_u - \frac{3}{2} \mass_d^\dag 
\mass_d + \fa - 
    \frac{17}{20} \tilde g_1^2 - \frac{9}{4} g_2^2 - 8g_3^2 ) 
\end{equation}
\begin{equation}\label{RGEYd}    
16 \pi^2 \, \,  \beta_{\mass_d} = 
\mass_d (\frac{3}{2} \mass_d^\dag \mass_d - \frac{3}{2} \mass_u^\dag 
\mass_u +  \fa -
\frac{1}{4}\tilde g_1^2 - \frac{9}{4}g_2^2 - 8 g_3^2 ) 
\end{equation}
\begin{equation}\label{RGEYnu}
 16 \pi^2 \, \,  \beta_{\mass_{\nu}} =   \mass_{\nu} (
\frac{3}{2}\mass_{\nu}^\dag \mass_{\nu}- \frac{3}{2}
\mass_e^\dag \mass_e + \fa - \frac{9}{20} \tilde g_1^2 - \frac{9}{4} g_2^2 ) 
\end{equation}
\begin{equation}\label{RGEYe}
 16 \pi^2 \, \,  \beta_{\mass_e} = \mass_e (
\frac{3}{2}\mass_e^\dag \mass_e- \frac{3}{2}
\mass_{\nu}^\dag \mass_{\nu}  +\fa  -\frac{9}{4} \tilde g_1^2 - \frac{9}{4} g_2^2) ,
\end{equation}
with $\fa$ as in \eqref{abcde} and where we use the notation 
$$ \tilde g_1^2= \frac{5}{3} g_1^2. $$
The RGE for the Majorana mass terms has beta function
\begin{equation}\label{RGEM}
 16 \pi^2 \, \,  \beta_{M} = 
\mass_\nu \mass_\nu^\dag M + M (\mass_\nu \mass_\nu^\dag)^T 
\end{equation}
and the one for the Higgs self coupling $\lambda$ is given by
\begin{equation}\label{RGElambda}
 16 \pi^2 \, \,  \beta_{\lambda} = 6 \lambda^2 - 3\lambda (3 g_2^2 + g_1^2) + 
 3 g_2^4 + \frac{3}{2} (g_1^2 + g_2^2)^2 + 4\lambda \fa -  8 \fb 
\end{equation} 
where the terms $\fa$ and $\fb$ are as in \eqref{abcde}.

Notice how, at one loop order, the equations for the three coupling constants
decouple from the other parameters. This is no longer true at two loops in the MSM,
see \cite{ACKPRW}. As explained in \cite{AKLRS}, these equations can be treated
as a series of equations for effective field theories between the different see-saw
scales and integrated numerically. As shown in \cite{MaKol}, there is a sensitive
dependence on the initial conditions at unification.

\medskip
\subsection{Approximations of RGEs}\label{approxRGE}

We recall here some useful approximations to the RGEs above
and some facts about the running of the solutions, with boundary
conditions at unification compatible with the constraints of the CCM
model recalled above. This analysis was already in \cite{CCM} but
we recall it here for direct comparison, later, with the modifications
due to the interaction with the gravitational terms.

The main idea is that, in first approximation, the top quark Yukawa
parameter is the dominant term in the equations \eqref{RGEYu},
\eqref{RGEYd}, and \eqref{RGEYe}. In the MSM case, it would
also be dominant over all the terms in \eqref{RGEYnu}, but in the
$\nu$MSM, because these are coupled to the Majorana masses
\eqref{RGEM}, the Yukawa coupling for the $\tau$ neutrino can 
also be of comparable magnitude.

If one neglects this additional term (that is, considers the RG
equations of MSM) then in the running of the Higgs self-coupling 
parameter $\lambda_t$ one can also neglect all terms except the 
coupling constants $g_i$ and the Yukawa parameter of the
top quark $y_t$. This gives equations
\begin{equation}\label{betalambda}
\beta_\lambda = \frac{1}{16 \pi^2} \left( 24 \lambda^2 + 12 \lambda y^2 - 9 \lambda 
(g_2^2 + \frac{1}{3} g_1^2) - 6 y^4 +\frac{9}{8} g_2^4 + \frac{3}{8} g_1^4 + \frac{3}{4} g_2^2 g_1^2\right)
\end{equation}
where the Yukawa parameter for the top quark has
\begin{equation}\label{betatop}
\beta_y  = \frac{1}{16 \pi^2} \left( \frac{9}{2} y^3 - 8 g_3^2 y - \frac{9}{4} g_2^2 y - \frac{17}{12} g_1^2 y \right).
\end{equation}

In \cite{CCM} the problem of the existence of an additional term
$y_{\tau}$ that contributes to the running is dealt with by 
modifying the boundary condition
at unification for the top Yukawa parameter. This is dictated, in the NCG model,
by the quadratic mass relation \eqref{massrelU}, which gives 
$$ \sum_{\sigma \in {\rm generations}} (y_\nu^\sigma)^2 +(y_e^\sigma)^2 + 3 (y_u^\sigma)^2 
+ 3 (y_d^\sigma)^2 = 4 g^2. $$
Of these, the dominant terms become
\begin{equation}\label{massrelterms}
 x_t^2 + 3\, y_t^2 = 4 g^2, 
\end{equation} 
where $x_t$ is the Yukawa coupling $y_\nu^{\sigma=3}(t)$ of the $\tau$ neutrino
(that is, for the generation index $\sigma=3$) and $y_t = y_u^{\sigma=3}(t)$ is
the top Yukawa coupling.  This gives a low evergy value $y_0$ for $y_t$ at
$y_0\sim 1.04$ as in \cite{CCM}, computed assuming a value of the coupling
constants at unification of $g\sim 0.517$ and setting the unification scale at
$\Lambda_{\rm unif}= 10^{17}$~GeV.  

\medskip
\subsection{Coupling constants equations without gravitational corrections}

\begin{figure}
\begin{center}
\includegraphics[scale=0.85]{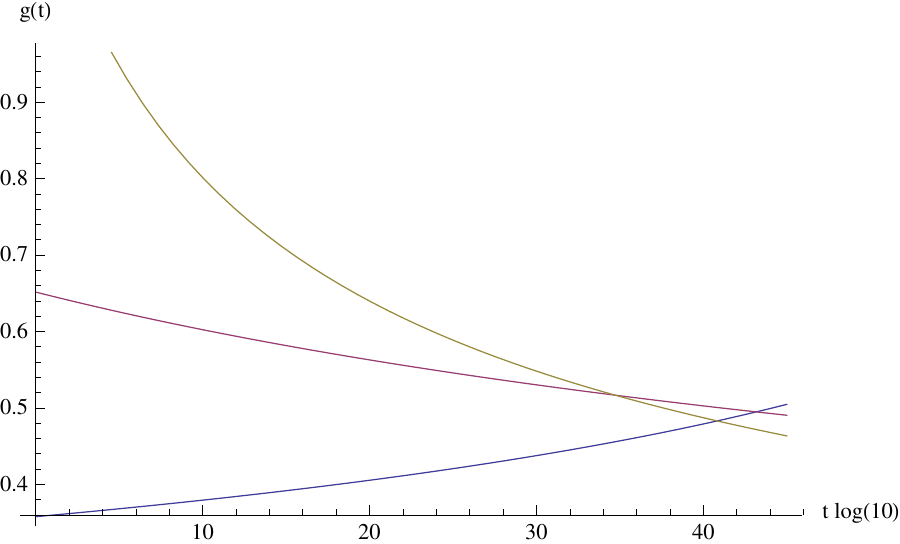}
\caption{Running of the coupling constants without anomalous dimensions (with ``unification" at around $10^{17}$ GeV). \label{grunFig}}
\end{center}
\end{figure}

The RGEs for the coupling constants \eqref{betaSMg} can be solved
exactly. In fact, an ODE of the
form 
\begin{equation}\label{yprimey3}
 u^\prime(t) = A\, u(t)^3, \ \ \ u(0)= B, 
\end{equation} 
has exact nontrivial solutions
\begin{equation}\label{yprimey3sol}
 u(t) = \pm \frac{1}{\sqrt{\frac{1-2 AB^2 t}{B^2}}}. 
\end{equation} 
At $\Lambda=M_Z$, the values of the coupling constants 
\begin{equation}\label{gMZ}
 g_1(0) = 0.3575, \ \  \ \	g_2(0) = 0.6514,	\ \ \ \
g_3(0) = 1.221 
\end{equation}
determine the values of $B$ in the above equation, while $A$ is determined
by the beta functions of \eqref{betaSMg}. This gives a running as illustrated in 
Figure \ref{grunFig}. One obtains the usual picture where, at high energies, 
the coupling constants do not quite meet, but form a triangle.

\medskip
\subsection{Top Yukawa coupling without gravitational corrections}

\begin{figure}
\begin{center}
\includegraphics[scale=0.85]{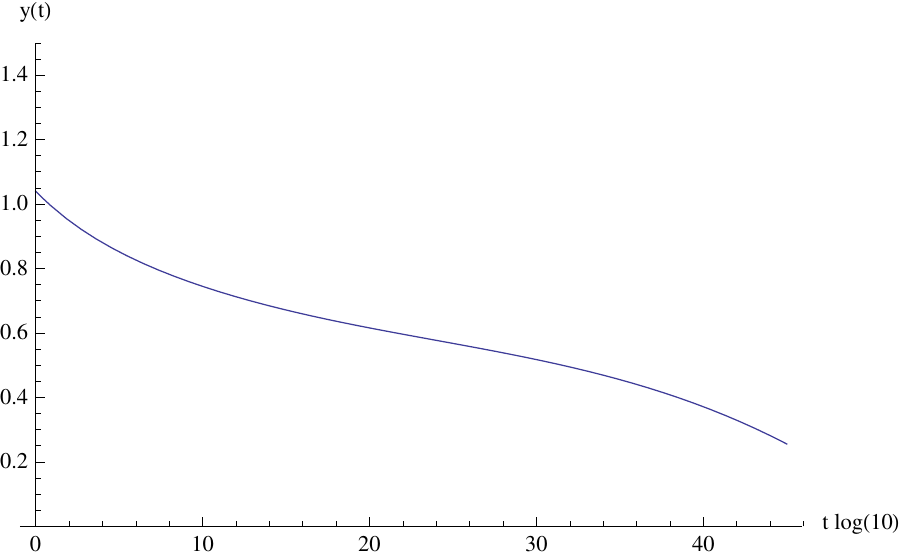}
\caption{Running of the top Yukawa coupling without anomalous dimensions. \label{yrunFig}}
\end{center}
\end{figure}

With the solutions \eqref{yprimey3sol} to the running \eqref{betaSMg} 
of the coupling constants $g_i(t)$, one can then solve numerically the equation
for the running of the top Yukawa parameter, with beta function \eqref{betatop},
using the boundary conditions at unification as in \cite{CCM}, as described above.
One obtains a running for $y(t)$ as in Figure \ref{yrunFig}. The running of the
Higgs quartic coupling can be analyzed in a similar way, by numerically solving
the equation with beta function \eqref{betatop}, see \cite{CCM} and \cite{CoMa}.

\medskip
\subsection{The running of the gravitational terms in the NCG models}

The RGE analysis of the gravitational terms carried out in \cite{CCM} is
based on the running of the conformal gravity term derived in \cite{Avra},
while estimates are given on the parameter $f_2$ of the model,
based on the value of the Newton constant at ordinary scales.
In particular, the running of the conformal gravity term with the Weyl curvature
$$ \alpha_0 \int C_{\mu \nu \rho \sigma} \, C^{\mu \nu \rho \sigma} $$
was analyzed in \cite{CCM} using as beta function (see equations (4.49) and (4.71)
of \cite{Avra})
$$ \beta_\eta = \frac{-1}{(4\pi)^2} \frac{133}{10} \eta^2, $$
with $\alpha_0 = \frac{1}{2\eta}$. It is shown in \cite{CCM} that
the running $\eta(t)$ does not change the value of $\eta$ by more
than a single order of magnitude between unification scale and 
infrared energies.
A different type of RGE analysis was done in \cite{MaPie}, where the
expressions \eqref{GeffU} and \eqref{gammaeffU} for effective gravitational
and effective cosmological constants of the model were run, with
the renormalization group equations of the $\nu$MSM model, between
unification and electroweak scales, to construct models of the very early
universe with variable gravitational and cosmological constants.

\smallskip

In the present paper, we propose a different use of the renormalization
group analysis, based on introducing correction terms to the RGEs of the
$\nu$MSM model coming from the coupling of matter to gravity, through
the method of Wetterich's functional renormalization group equations (FRGE) of
\cite{Wett}, as in \cite{ShaWett}.

\section{Renormalization Group Equations and the Spectral Action}\label{RGEsec}

As recalled in the previous section, generally in NCG models 
the RGE for the particle physics sector are imported from the
corresponding particle physics model (MSM, as in \cite{ACKPRW}, 
or an extension with right handed neutrinos and Majorana mass terms, 
see \cite{AKLRS}). The fact that, in the NCG models, matter is coupled
to gravity only enters through the constraints on the boundary conditions
at unification described above, but not in the form of the RGEs themselves.
We propose here a different RGE analysis, where an effect of the
gravitational term is manifest also at the level of the RGEs. 

\medskip
\subsection{Wetterich's FRGE and the spectral action}

There is a general method, developed in \cite{Wett}, to
derive renormalization group equations from an effective
action. 
In terms of the spectral action functional, one can regard
asymptotic expansion for large $\Lambda$ as the large cutoff 
scale limit  of an effective action, somewhat like in \cite{Wett}. 
Namely, we have a cutoff scale in the spectral action, if we think 
of the test function $f$ as a smooth approximation to a cutoff
function. We can then interpret the spectral action $\Tr(f(D/\Lambda))$
itself as the ``flowing action" in the sense of Wetterich, with
the dependence on $\Lambda$ in $f(D/\Lambda)$ moving the
cutoff scale on $D$. In the limit for $\Lambda \to \infty$, we
can, to very high precision (see \cite{CC3}) 
replace the spectral action by its asymptotic expansion,
which recovers the classical action for gravity coupled to
matter. In the notation of the Wetterich approach we would
then denote by $\cS_k = \Tr(f(D/\Lambda))$ the
spectral action where $f(x)$ is an even smooth approximation of
a cutoff function with cutoff at $x= k>0$.

\smallskip

One could consider a functional renormalization group equation,
in the sense of Wetterich, associated to the spectral action functional, in the form
\begin{equation}\label{RGESA}
\partial_t \cS_k = \frac{1}{2} \Tr \left( \left( \frac{\delta \cS_k}{\delta A \delta A} + \cR_k \right)^{-1}
\partial_t \cR_k \right),
\end{equation}
where $t\sim \log \Lambda$ and $k$ is the cutoff scale for the test function $f$,
the $A=(A_i)$ are the boson fields, viewed as the independent fluctuations of
the Dirac operator in the manifolds and non-commutative directions, and $\cR_k$ is
a tensorial cutoff as in \cite{Wett}. A detailed discussion of how to derive RGEs 
from the spectral action with this method, the effect of adding the fermionic term 
of the spectral action as in \eqref{SAfermions}, and the relation between the
RGEs obtained in this way to the RGEs of the $\nu$MSM of \cite{AKLRS} will
be given elsewhere.

\smallskip

For our purposes here, it suffices to observe that, for sufficiently
large $k$, the non-perturbative spectral action is very well
approximated by an action of the from 
\begin{equation}\label{FVphiR}
\cS[g,\phi] = \int_X \left( \cV(\phi^2) - \cF(\phi^2) R + \frac{1}{2} (\partial \phi)^2 \right)\, \sqrt{g}\, d^4x,
\end{equation}
for suitable functions $\cV$ and $\cF$.
Strictly speaking, this is true in the case of sufficiently regular 
(homogeneous and isotropic) geometries, for which the 
spectral action can be computed non-perturbatively, in terms of Poisson
summation formulae applied to the spectrum of the Dirac operator,
as shown in \cite{CC3}, see also \cite{CaMaTeh}, \cite{MaPieTeh1}, 
\cite{MaPieTeh2}, and \cite{Teh}. In such cases, the functions $\cV$
and $\cF$ take the form of a slow-roll potential for the Higgs field, or
for more general scalar fluctuations of the Dirac operator
of the form $D^2 \mapsto D^2 +\phi^2$. These potentials have
been studied as possible models of inflationary cosmology.
A derivation of the non-perturbative spectral action and the
slow-roll potential in terms of heat-kernel techniques is also
given in \cite{CaMaTeh}. 

\smallskip

For more general geometries, in adopting the form \eqref{FVphiR},
as opposed to the more general \eqref{largeLambda},
we are neglecting the additional gravitational term
given by the Weyl curvature. While these can produce interesting
effects that distinguish the NCG model of gravity from ordinary
general relativity (see \cite{NeOcSa}, \cite{NeOcSa2}
for a discussion of some interesting cases), 
the term is subdominant to the Einstein--Hilbert term
at unification scale \cite{MaPie}. The argument  from \cite{CCM} recalled above 
about the running of this term shows that it changes by at most an order of 
magnitude at lower scales, so we can assume that it remains subdominant and
neglect it in first approximation.

\medskip
\subsection{RGE with gravitational corrections from the spectral action}

The Wetterich method of FRGE 
was used in \cite{DouPe} and \cite{Reut} to study the 
running of gravitational and cosmological constant for the Einstein--Hilbert 
action with a cosmological term. The existence of an attractive UV fixed point 
was shown in \cite{Souma} and further studied in \cite{LauReut}. This
result was further generalized in \cite{PePe} to the case of a theory
with a scalar field non-minimally coupled to gravity via a pairing with
the scalar curvature of the form \eqref{FVphiR} as above, 
where the values $\cV(0)$ and $\cF(0)$ depend on the cosmological and gravitational
constant, so that \eqref{FVphiR} also contains the usual Einstein--Hilbert and 
cosmological terms. The functions $\cV$ and $\cF$ are assumed to be polynomial in \cite{PePe},
with $\cV$ at most quartic and $\cF$ at most quadratic, although most of the
results of \cite{PePe} (see the generalization given in \cite{NaPe}) work for more 
general real analytic $\cV$ and $\cF$, which include the slow roll
case. A detailed analysis of the critical surface 
of flow trajectories of the RGE approaching the UV fixed
point is carried out in \cite{PePe}.

The explicit form of the RGE for an action of the form \eqref{FVphiR}
were computed explicitly in \cite{NaPe}, \cite{PePe}. By the discussion
above, they can serve as a good model of RGE for the spectral action functional.
After making the same approximations as discussed in \S \ref{approxRGE},
the resulting equations become of the form used in \cite{ShaWett}, which
we will discuss in detail in \S \ref{RGEgravSec} below.

\medskip
\subsection{RGE flow as a running of the noncommutative geometry}\label{NCGrunSec}

In terms of the geometry of the model, the renormalization group
flow has a natural interpretation in terms of a running of the
finite noncommutative geometry $F=(\cA_F, \cH_F, D_F)$.
In fact, as shown in \cite{CCM} (see also Chapter 1 of \cite{CoMa}),
the Dirac operator $D_F$ of the finite geometry encodes all the
Yukawa coupling matrices $Y$ of \eqref{RGEYu}--\eqref{RGEYe},
as well as the Majorana mass term matrix \eqref{RGEM}. Moreover,
the data of the Dirac operator $D_F$ on the finite geometry
are described by a moduli space of the form $\cM_F=\cG_\ell \times \cG_q$,
with the quark sector given by
$$ \cG_q= (U(3)\times U(3))\backslash (\GL_3(\C)\times \GL_3(\C))/U(3), $$
and the lepton sector 
$$ \cG_\ell =  (U(3)\times U(3))\backslash (\GL_3(\C)\times \GL_3(\C)\times \cS_3)/U(3), $$
with $\cS_3$ the space of symmetric complex $3\times 3$-matrices.
(For a more general discussion of moduli spaces of finite spectral triples
see also \cite{Cac}.)
This space describes the {\em bare parameters} that enter the 
spectral action, in the form of the Dirac operator $D_A$, with
$D=\dirac_X \otimes 1 \oplus \gamma_5 \otimes D_F$ the
Dirac operator of the product geometry and $A$ the inner
fluctuations.
Thus, the one-loop RGE flow can be interpreted as a flow on $\cM_F$, for
given solutions $g_i(t)$ of the RGE for the coupling constants. 

This can be better interpreted as a flow on a fibration $\tilde \cM_F$
over a 3-dimensional base space $\cB_F$, with fiber the moduli space $\cM_F$, 
where the base space $\cB_F$ corresponds to the possible
values of the coupling constants $(g_1,g_2,g_3)$. The renormalization group
flow then becomes a flow in $\tilde \cM_F$. At one-loop level, where
the equation for the coupling constants uncouples from the equations for
the Yukawa parameters and Majorana mass terms, one can see the solutions
to the RGE flow for the coupling constants as defining a curve in the base space
$\cB_F$ and the RGE for the remaining parameters as defining a curve in 
$\tilde \cM_F$ that covers the curve in $\cB_F$.

\section{Renormalization group equations with anomalous dimensions}\label{RGEgravSec}

Asymptotic safety, as formulated in \cite{Wein}, is a generalization of
the notion of renormalizability, which consists of the requirement that
the coupling constants lie on the critical surface of the UV fixed point.
This allows for renormalization group analysis of gravitational terms
coupled to matter.
One of the main consequences of the asymptotic safety scenario is that the RGE for
the matter sector acquires correction terms coming from the gravitational
parameters in the general form
\begin{equation}\label{betagravRGE}
\partial_t x_j = \beta_j^{\rm SM} + \beta_j^{\rm grav},
\end{equation}
where $x_j$ are the running parameters, $t=\log \Lambda$, and
$\beta_j^{\rm SM}$ is the Standard Model beta function for $x_j$
and $\beta_j^{\rm grav}$ is the gravitational correction. The latter
is of the form
\begin{equation}\label{betagrav}
\beta_j^{\rm grav} = \frac{a_j}{8 \pi} \frac{\Lambda^2}{M_P^2(\Lambda)}\, x_j,
\end{equation}
where the $a_j$ are the {\em anomalous dimensions}.

The scale dependence of the Newton constant is given by (see \cite{Reut})
\begin{equation}\label{NewtonLambda}
M_P^2(\Lambda)= M_P^2  + 2 \rho_0 \Lambda^2,
\end{equation}
where the parameter $\rho_0$ that expresses this scale dependence is
estimated to have a value of $\rho_0\sim 0.024$, see \cite{Reut} and 
\cite{PePe}, \cite{NaPe}.

We now consider the presence of the
correction terms to the RGE coming from the gravitational sector, via
the additional terms \eqref{betagrav} in the  beta functions \eqref{betagravRGE}.

\medskip
\subsection{Anomalous dimensions in the gauge sector} 
\begin{figure}
\begin{center}
\includegraphics[scale=0.85]{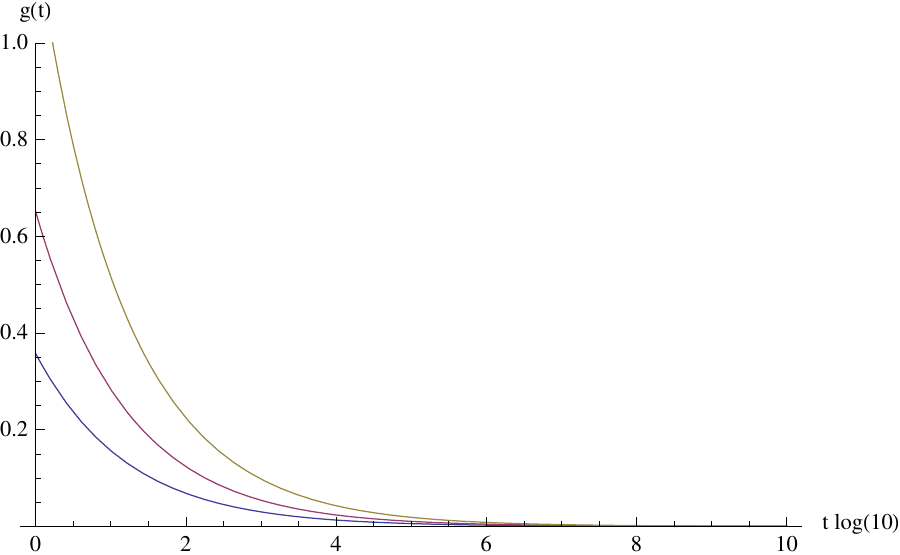}
\caption{Running of the coupling constants with anomalous dimensions. \label{grun2Fig}}
\end{center}
\end{figure}

As in \cite{ShaWett}, we assume that the $a_j$ for the couping constants
are all equal and depending only on their common value at unification
$a_1=a_2=a_3=a_g$, with $|a_g| \sim 1$ and a negative sign. 
Thus, instead of an equation of the form \eqref{yprimey3} for the running of
the coupling constants, we obtain a modified equation of the form
\begin{equation}\label{yprimeyy3}
u^\prime(t) = -a \, u(t) + A \, u(t)^3, \ \ \ \ u(0)=B ,
\end{equation}
which again can be solved exactly and has (in addition to the trivial solution)
solutions of the form
\begin{equation}\label{yprimeyy3sol}
u(t) = \frac{\pm \sqrt{a}}{\sqrt{ A + \exp\left( 2a \left( t+\displaystyle{ \frac{\log(-A + \frac{a}{B^2})}{2a}  }\right) \right)}  } .
\end{equation}
Again imposing conditions \eqref{gMZ} to fix $B$, and setting 
$$ a = \frac{|a_g|}{16 \pi \rho_0} \sim \frac{1}{16 \pi \rho_0} $$
with $A$ determined by the beta
functions \eqref{betaSMg}, with $\rho_0\sim 0.024$ as above.
We find that all the couplings run rapidly to
zero at high energies, as illustrated in Figure \ref{grun2Fig}.

\medskip
\subsection{Anomalous dimension for the top Yukawa coupling}
We have seen above that, at high energies, the coupling constant terms 
are then subdominant in the running for
$y$ and $\lambda$ for sufficiently large $\Lambda$. For the top Yukawa coupling, 
the analysis of \cite{ShaWett} shows that the anomalous dimension satisfies $a_y <0$. 
Thus, one obtains an RGE equation for $y_t$ of the simpler form
\begin{equation}\label{yRGE}
 \partial_t y_t = -\frac{|a_y|}{16 \pi \rho_0} y_t + \frac{1}{16 \pi^2} \frac{9}{2} y_t^3. 
\end{equation} 
This can be solved exactly, with solutions that are again of the general form
\eqref{yprimeyy3sol}, this time with parameters
\begin{equation}\label{aAy}
 a = \frac{|a_y|}{16 \pi \rho_0}, \ \ \ \  A= \frac{9}{32 \pi^2} . 
\end{equation} 
Again, the parameter is set to $\rho_0\sim 0.024$.
There is an infrared fixed point with 
$$  y_{IR}^2 = \frac{2\pi |a_y|}{9 \rho_0}. $$ 

We analyze the running of the top Yukawa coupling, in the presence
of anomalous dimensions, using \eqref{yRGE} with the assumption,
as discussed in \cite{ShaWett}, that $a_g\leq a_y <0$. 
The initial condition can be taken as in \cite{CCM}, with 
$y(0)=B=1.04$. We obtain a running where $y(t)$
rapidly decays to zero before reaching unification scale. One sees in this 
way the well known fact that the presence of 
the gravitational correction terms makes the Yukawa couplings 
asymptotically free. 

\begin{figure}
\begin{center}
\includegraphics[scale=0.85]{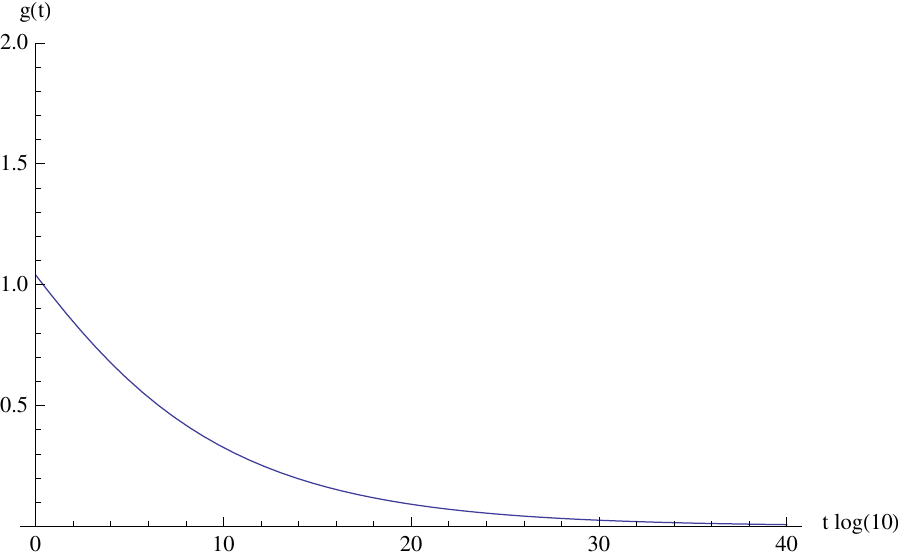}
\caption{Running of the top Yukawa coupling with anomalous dimension $|a_y|\sim 0.155$. 
\label{yrun2Fig}}
\end{center}
\end{figure}

\smallskip
\subsubsection{Anomalous dimension for the Higgs self-coupling}

The qualitative analysis of the running of the Higgs self-coupling $\lambda$
given in \cite{ShaWett} can be equally applied within the setting of the NCG model.
They use an estimate on the anomalous dimension $a_\lambda \sim 3.1$, as
suggested in \cite{PePe}, \cite{NaPe}. (We will discuss in \S\S \ref{HypSec} and \ref{HypSec2} 
some other natural choices of $a_\lambda$.) The effective action is of the form \eqref{FVphiR},
which, as we discussed in \S \ref{FVphiR} above is a very good approximation 
for the spectral action for $\Lambda$ running between the electroweak and
the unification scale. The qualitative analysis of  \cite{ShaWett} is based on
estimating, as above, that the top Yukawa coupling contribution to the beta function for
the Higgs self-coupling is dominant over the gauge contribution, which leads
to a further simplification of the equation to 
\begin{equation}\label{lambdaRGE}
\partial_t\lambda= \frac{a_\lambda}{16 \pi \rho_0} \lambda +
\frac{1}{16 \pi^2} ( 24 \lambda^2 + 12 \lambda y^2 - 6 y^4), 
\end{equation}
coupled to the running \eqref{yRGE} of the top Yukawa $y_t$. Here again $\rho_0  \sim 0.024$.

The equation \eqref{lambdaRGE} also gives an approximate 
fixed point for $\lambda$ determined by the equation
\begin{equation}\label{lambdayfixed}
 24 \lambda^2 + 12 \lambda y_{IR}^2 - 6 y_{IR}^4 + \frac{\pi a_\lambda}{\rho_0} \lambda =0, 
\end{equation} 
or equivalently, 
\begin{equation}\label{lambdaIR}
 24 \lambda^2 + \frac{\pi}{\rho_0} ( \frac{8}{3} |a_y| + a_\lambda) \lambda - \frac{4\pi^2}{27}
\frac{|a_y|^2}{\rho_0^2} =0. 
\end{equation}

\medskip
\subsection{Riccati equations and singular linear ODEs}

After substituting the explicit general form of the solution for the equation \eqref{yRGE}
for the function $y$, the equation \eqref{lambdaRGE} for the unknown function $\lambda$
becomes a nonlinear ODE of Riccati type, 
\begin{equation}\label{Riccati}
\lambda^\prime = q_0(t) + q_1(t) \lambda + q_2(t) \lambda^2, 
\end{equation}
in our case with
\begin{equation}\label{qit}
q_0(t)= \frac{-3 y^4(t)}{8\pi^2}, \ \ \  q_1(t)=\frac{a_\lambda}{16\pi \rho_0} + \frac{3 y^2(t)}{4 \pi^2}, 
\ \ \  q_2(t) = \frac{3}{2 \pi^2}.
\end{equation}
Riccati equations have the property that they
can be transformed into a {\em linear} second order ODE by a change of variables of the
form 
\begin{equation}\label{ulambda}
- \frac{u^\prime}{u}=\lambda\, q_2 .
\end{equation}
This gives a linear equation of the form
\begin{equation}\label{linueq}
u^{\prime\prime} - \left( q_1(t) + \frac{q_2^\prime(t)}{q_2(t)} \right)\, u^\prime + 
q_2(t) q_0(t)\, u =0.
\end{equation}

Using a general form
$$ y(t)^2 =\frac{a}{Ce^{2at}+A} $$
for the solutions of \eqref{yRGE}, with parameters $a$, $A$, and $C$ determined
in terms of the coefficients of \eqref{yRGE} and the initial condition, as discussed above,
we obtain a second order linear equation of the form
\begin{equation}\label{linearODE}
 \partial_t^2 u(t) - (\frac{a_\lambda}{16\pi\rho_0}+\frac{3}{4\pi^2}\frac{a}{Ce^{2at}+A})\partial_t u - \frac{9}{16\pi^4} (\frac{a}{Ce^{2at}+A})^2 u(t) =0 .
\end{equation}

After changing variables to $x=e^{2at}$, with $v(x)=v(e^{2at})=u(t)$, we can write the
above as
\begin{equation}\label{SingODEx}
(2ax)^2\partial_x^2 v(x) + (2a-\frac{a_\lambda}{16\pi\rho_0}-\frac{3}{4\pi^2}\frac{a}{Cx+A})(2ax\partial_x v(x)) - \frac{9}{16\pi^4}(\frac{a}{Cx+A})^2v(x)=0 .
\end{equation}

\medskip
\subsection{Polynomial solutions of singular linear ODEs}

We consider here special relations between the parameters
$a$, $C$, $A$, and $a_\lambda$ that lead to the existence of
special polynomial solutions of \eqref{SingODEx}.

Following \cite{Saad}, given a linear differential equation of the form
\begin{equation}\label{ODEsl}
y^{\prime\prime}=\lambda_0 y^\prime + s_0 y,
\end{equation}
where $\lambda_0$ and $s_0$ are $\cC^\infty(I)$ functions on some
interval of real numbers $I\subseteq \R$, one can recursively obtain
equations of a similar form for the higher derivatives,
\begin{equation}\label{ODEsln}
y^{(n+1)}=\lambda_{n-1}\, y^\prime + s_{n-1}\, y,
\end{equation}
where the functions $\lambda_n$ and $s_n$ are recursively
defined by
\begin{equation}\label{lnsn}
\begin{array}{rl}
\lambda_n = & \lambda_{n-1}^\prime + s_{n-1} +\lambda_0 \lambda_{n-1} \\[2mm]
s_n = & s_{n-1}^\prime + s_0 \lambda_{n-1}.
\end{array}
\end{equation}
Then the equation \eqref{ODEsl} has a polynomial solution of degree at most $n$ if 
$\lambda_n \lambda_{n-1}\neq 0$ and 
\begin{equation}\label{deltan0}
\delta_n:= \lambda_n \, s_{n-1} - \lambda_{n-1}\, s_n =0.
\end{equation}
One then has a solution of the form 
\begin{equation}\label{solyn}
y = \exp \left( -\int \frac{s_n}{\lambda_n}  \right) = 
\exp \left(- \int \frac{s_{n-1}}{\lambda_{n-1}}  \right),
\end{equation}
and a general solution of the form
\begin{equation}\label{solyngen}
y = \exp \left(- \int \frac{s_n}{\lambda_n}  \right) \left( C_1 + C_2 \int \exp\left( \int \lambda_0 + 
2 \frac{s_n}{\lambda_n}  \right) \right)
\end{equation}

The equation \eqref{SingODEx} can be put in the form 
\eqref{ODEsl} with 
\begin{equation}\label{l0}
\lambda_0 = 
\frac{-\frac{3a}{4\pi^2} + \left( 2a - \frac{a_\lambda}{16 \pi \rho_0}\right) (A+Cx)}{2ax (A+Cx)}
\end{equation}
\begin{equation}\label{s0}
s_0 = -\frac{9}{64 \pi^2 x^2 (A+Cx)^2}.
\end{equation}
The first step of the recursive equation \eqref{lnsn} gives
\begin{equation}\label{l1}
\begin{array}{rl}
\lambda_1 = & \frac{1}{(A + Cx)^2} \left(C^2 + 
\frac{C^2 a_\lambda^2}{1024 a^2 \pi^2 \rho_0^2} 
- \frac{C^2 a_\lambda}{16 a \pi \rho_0}\right) \\[3mm]
+ &  \frac{1}{(A + Cx)^2 x^2} \left(A^2 + \frac{9}{64  \pi^4}
-\frac{3A}{4  \pi^2} + \frac{A^2 a_\lambda^2}{1024 a^2  \pi^2 \rho_0^2}
+ \frac{3 A a_\lambda}{ 128 a \pi^3 \rho_0}
- \frac{A^2 a_\lambda}{ 16 a \pi \rho_0}
- \frac{9}{64 \pi^2}\right) \\[3mm]
+ & \frac{1}{(A+Cx) x^2}\left( -A + \frac{3}{8 \pi^2}
+ \frac{A a_\lambda}{32 a  \pi \rho_0} \right) \\[3mm]
+ & \frac{1}{(A+Cx)^2 x}\left( 2AC -\frac{3C}{4 \pi^2}
+ \frac{A C a_\lambda^2}{512 a^2  \pi^2 \rho_0^2}
+\frac{3C a_\lambda}{128 a \pi^3 \rho_0}
-\frac{AC a_\lambda}{ 8 a  \pi \rho_0}\right),
\end{array}
\end{equation}
\begin{equation}\label{s1}
\begin{array}{rl}
s_1 = & \frac{1}{(A+Cx)^3 x^3} \left( \frac{27}{512 \pi^4} + \frac{9A}{64 \pi^2} +
\frac{9 A a_\lambda}{2048 a \pi^3 \rho_0} \right) \\[3mm]
+ & \frac{1}{(A+Cx)^3 x^2} \left( \frac{27 C}{64 \pi^2} + \frac{9 C a_\lambda}{2048 a \pi^3 \rho_0}
\right) .
\end{array}
\end{equation}
Thus, at this first step one obtains
$$ \delta_1 = \frac{1}{x^4(A+C x)^5} \left( \frac{81 A}{4096 \pi ^4}+\frac{27 A^2}{512 \pi ^4}
-\frac{9 A^3}{64 \pi ^2}+\frac{9 A^3 a_\lambda}{2048 a \pi ^3 \rho_0} \right) + $$
$$ \frac{1}{x^3 (A+C x)^5} \left( 
\frac{81 C}{4096 \pi ^4}+\frac{27 A C}{128 \pi ^4}-\frac{27 A^2 C}{32 \pi ^2}+\frac{27 A^2 C a_\lambda}{1024 a \pi ^3 \rho_0} \right) + $$
$$ \frac{1}{x^2 (A+C x)^5}\left( 
\frac{81 C^2}{512 \pi ^4}-\frac{81 A C^2}{64 \pi ^2}+\frac{27 A C^2 a_\lambda}{1024 a \pi ^3 \rho_0} \right) + $$
$$ \frac{1}{x (A+C x)^5}\left( -\frac{9 C^3}{16 \pi ^2}+\frac{9 C^3 a_\lambda}{512 a \pi ^3 \rho_0} \right) . $$
The vanishing $\delta_1\equiv 0$ implies that the parameters should satisfy
either $C=0$ and $A=0$ or else $C=0$ and 
\begin{equation}\label{alambdaVal}
 a_\lambda = \frac{(-9 - 24 A + 64 A^2 \pi^2) a \rho_0}{2 A^2 \pi}. 
\end{equation} 
Since both these solutions have $C=0$ they are ruled out for physical reasons.
We nonetheless look more closely at the second case. 
The linear solution in this case is obtained by replacing these
values of $C$ and $a_\lambda$ in \eqref{l0} and \eqref{s0}, so that one gets
$$ \frac{s_0}{\lambda_0}|_{C=0, a_\lambda = \frac{(-9 - 24 A + 64 A^2 \pi^2) a \rho_0}{2 A^2 \pi}} = - \frac{1}{x} . $$
A solution is then given by $y=\exp(\int s_0/\lambda_0)=x$ and the general
solution by 
$$ y= C_1 x + C_2 x^{\alpha +2}, $$
where
$$ \alpha = 2 + \frac{9}{64 A^2 \pi^2} , $$
with the latter obtained from the fact that 
$$ \lambda_0 |_{C=0, a_\lambda = \frac{(-9 - 24 A + 64 A^2 \pi^2) a \rho_0}{2 A^2 \pi}} = \frac{9}{64 A^2 \pi^2 x}. $$ 

\medskip

At the second order, the explicit expressions for $\lambda_2$ and $s_2$ are
more involved, but one encounters essentially the same situation. Namely,
the condition for the vanishing $\delta_2\equiv 0$ is given by a system of
five equations in the parameters. 
These equations have solutions
$$ C =0,  \ \ \ \ \  a_\lambda = \frac{(-9-24 A+64 A^2 \pi ^2) a \rho_0}{2 A^2 \pi} $$
$$ C =0,  \ \ \ \ \  a_\lambda =-\frac{3a \rho_0(3  +16 A)}{4 A^2 \pi}, $$
which again are not compatible with having $C\neq 0$ in the solution of the RGE
equation for the top Yukawa coupling $y$. 

\medskip

One can see a similar situation for the
case of $\lambda_3$ and $s_3$, where one has several more polynomial solutions,
but all of them again under the condition that $C=0$. In this case one obtains a
system of seven equations for
the parameters that arise from imposing $\delta_3=0$. 
The possible solutions to these equations with $A$ positive real
are the following:
\begin{equation}\label{parDeg3}
\begin{array}{lll}
 & a_\lambda = \displaystyle{\frac{a \left(-9-24 A+64 A^2 \pi ^2\right) r}{2 A^2 \pi }} & C=0  \\[2mm]
 & a_\lambda= \displaystyle{-\frac{a \left(3+24 A+64 A^2 \pi ^2\right) r}{2 A^2 \pi }}  & C=0 \\[2mm]
 & a_\lambda = \displaystyle{-\frac{3a \rho_0 (3  +16 A )}{4 A^2 \pi }} & C=0 \\[2mm]
 A =\displaystyle{\frac{24+\sqrt{576+4608 \pi ^2}}{256 \pi ^2}}  & a_\lambda =  -32 a \pi  \rho_0   & C=0 \\[2mm]
 A= \displaystyle{\frac{48+\sqrt{2304+4608 \pi ^2}}{256 \pi ^2}} & a_\lambda =  -32 a \pi  \rho_0   &   C=0 
\end{array}
\end{equation}

Notice how, at least up to degree three, the only condition that is compatible
with having a positive anomalous dimension $a_\lambda >0$ is the relation
\eqref{alambdaVal} (though not for all values of $A$).

\medskip
\subsection{Higgs self-coupling and hypergeometric functions}\label{HypSec}

We have seen above that, in the non-physical case with $C=0$ and with
the anomalous dimension $a_\lambda$ of the form \eqref{alambdaVal},
the equation \eqref{SingODEx} admits linear solutions $y= C_1 x$. Here
we drop the unphysical condition $C=0$, but we keep the same constraint
on $a_\lambda$ given \eqref{alambdaVal}, and we show that then the
resulting equation \eqref{SingODEx} can be integrated in closed form
and has a general solution that can be expressed in terms of 
hypergeometric functions.

Notice first that, after imposing the condition  \eqref{alambdaVal}, the
equation \eqref{SingODEx} can be written in the form \eqref{ODEsl},
with
\begin{equation}\label{newl0s0}
\begin{array}{rl}
\lambda_0 = & \displaystyle{ -\frac{1}{(A+C x)}
\left(\frac{9 C}{64 A^2 \pi ^2 }+\frac{3 C}{8 A \pi ^2 }+\frac{9}{64 A \pi ^2 x }\right) }\\[4mm]
s_0 = & - \displaystyle{ \frac{9}{64 \pi ^2 x^2 (A+C x)^2}}
\end{array}
\end{equation}

We introduce the following auxiliary functions:
\begin{equation}\label{f1}
f_1(x)=  (A+C x)^{\alpha-3}
\end{equation}
with
\begin{equation}\label{theta}
\theta= \frac{\sqrt{9+4 \pi ^2 \left(-9+4 A \left(3+4 A \pi ^2\right)\right)}}{16 A \pi ^2}
\end{equation}
and
\begin{equation}\label{alphax}
\alpha = \frac{3}{2}+\frac{3}{16 A \pi ^2}+\theta,
\end{equation}
\begin{equation}\label{f2}
f_2(x)=  x^{\eta} \left(\frac{C x}{A}\right)^{1-\eta}  \ \ \  \text{ with } \ \ \eta = \frac{9}{64\, A^2 \pi ^2}.
\end{equation}
We also set
\begin{equation}\label{betax}
\beta = \alpha -\eta -1 - \frac{3}{8A \pi^2},
\end{equation}
\begin{equation}\label{deltax}
\gamma(x)= -\frac{C x}{A}.
\end{equation}
Then the equation \eqref{ODEsl} with \eqref{newl0s0} has general solution of the form
\begin{equation}\label{Solution}
\begin{array}{rl}
v(x)= & C_1 \, \,  f_2(x)\, f_1(x) \,\,\, {}_2F_1\left( \alpha, \beta, 2-\eta, \gamma(x) \right) \\[3mm] + &
 C_2 \, \,  x^\eta \, f_1(x) \,\,\, {}_2F_1\left( \alpha-(2+\frac{3}{8A\pi^2}) , \beta+(2+\frac{8A}{3})\eta ,\eta , \gamma(x) \right),
\end{array} 
\end{equation}
where ${}_2 F_1(a,b,c,z)$ is the Gauss hypergeometric function.

\begin{figure}
\begin{center}
\includegraphics[scale=0.85]{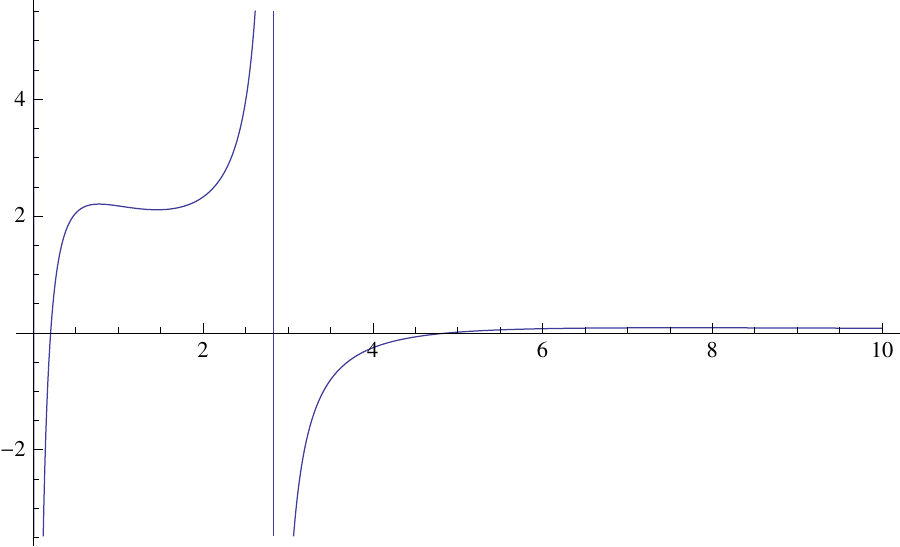}
\caption{Solutions of the form \eqref{lambdaSol} with $v(x)$ as in \eqref{Solution}
can exhibit complicated and singular behavior, depending on the values
of the parameters $C_1$, $C_2$, $A$, $C$, $a$. 
\label{lambdajumpsFig}}
\end{center}
\end{figure}

The corresponding solutions for the original Riccati equation \eqref{lambdaRGE}
that gives the renormalization group flow of the Higgs self coupling $\lambda$ are
then of the form 
\begin{equation}\label{lambdaSol}
\lambda(t) = -  \frac{2\pi^2}{3} \frac{u^\prime(t)}{u(t)}, \ \ \ \text{ with } \ \ \  u(t)=v(e^{2 at}).
\end{equation}
As was already observed in \cite{ShaWett}, these solutions, for varying choices
of the parameters $C_1$, $C_2$, $A$, $C$, $a$, can exhibit very complicated and
singular behavior (see an example in Figure \ref{lambdajumpsFig}). 

\begin{figure}
\begin{center}
\includegraphics[scale=0.85]{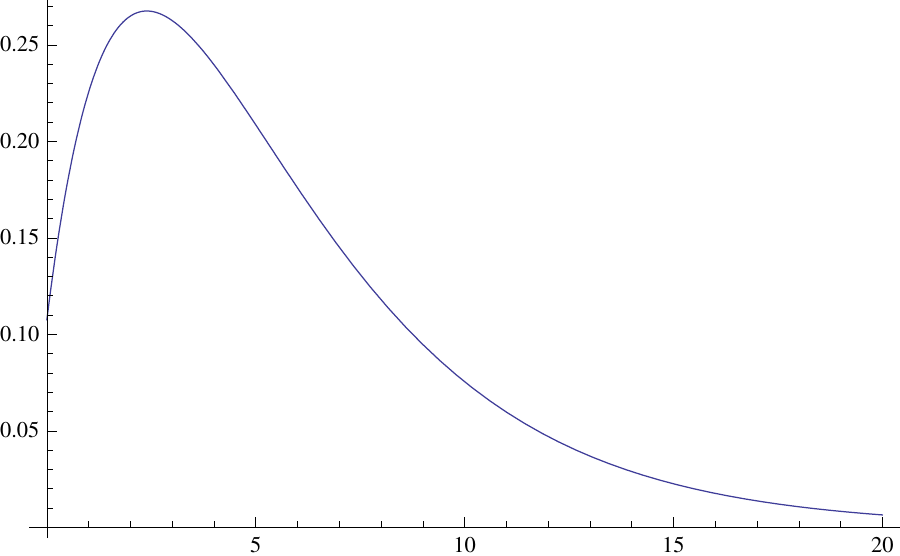}
\caption{A solution with $\lambda(t)$ continuous and
positive for $t\geq 0$. 
\label{lambdatpositiveFig}}
\end{center}
\end{figure}

Thus, one needs to ensure the existence of particular solutions 
that will be compatible with physical assumptions on $\lambda$. 
So at least one needs a solution for which $\lambda(t)$ is positive
and does not develop singularities for all $t>0$. This is, in fact,
easy to achieve. For example, setting $A=1$ and $C\geq 2$,
and $a=|a_y|/(16\pi \rho_0)$, with $|a_y|= 0.155$
and $\rho_0= 0.024$, gives a solution of the form as in Figure
\ref{lambdatpositiveFig}.

However, solutions obtained in this way are still not entirely
satisfactory from the physical point of view for two main reasons.
The first is that we are using the relation \eqref{alambdaVal}
that expresses $a_\lambda$ as a function of $A$, $a$ and $\rho_0$.
While the example $A=1$ above gives a plausible value of
the anomalous dimension $a_\lambda$, the value $A=\frac{9}{32 \pi^2}$
considered in \eqref{aAy} would give $a_\lambda$ the wrong sign.
The other reason is that a solution like the one illustrated in 
Figure \ref{lambdatpositiveFig}, while well behaved in the range $t\geq 0$, 
is still plagued by infrared divergences, although these can be pushed
very far back in the $t<0$ range.

\medskip
\subsection{Other special solutions of the Higgs self-coupling ODE}\label{HypSec2}

The form of the singular linear ODE \eqref{SingODEx} also suggests as a 
possible natural choice of $a_\lambda$ the value
\begin{equation}\label{alambda2}
a_\lambda = 32 \pi \rho_0 a = 2 \, |a_y|.
\end{equation}
For this value the equation \eqref{SingODEx} simplifies to the form
\begin{equation}\label{SingODEx2}
(2ax)^2\partial_x^2 v(x) + -\frac{3}{4\pi^2}\frac{a}{Cx+A} 2ax\partial_x v(x) - \frac{9}{16\pi^4}(\frac{a}{Cx+A})^2v(x)=0 ,
\end{equation}
or equivalently to an equation of the form \eqref{ODEsl},
with
\begin{equation}\label{l0s0aa}
\begin{array}{rl}
\lambda_0 = & - \displaystyle{ \frac{3}{8 \pi ^2 x (A+C x)}} \\[4mm]
s_0 = & - \displaystyle{ \frac{9}{64 \pi ^2 x^2 (A+C x)^2}} = - \pi^2 \lambda_0^2 
\end{array}
\end{equation}
In this case, again, we have an explicit family of solutions in terms of hypergeometric
functions. We introduce the following notation.
\begin{equation}\label{h1h2x}
h_1(x)=  (x (A+ Cx))^{\upsilon}, \ \ \ h_2(x)= \left(\frac{Cx}{A}\right)^{-2\tilde\theta}
\end{equation}
with 
\begin{equation}\label{upsilon}
\upsilon = \frac{1}{2}-\frac{3}{16 A \pi^2}+\theta
\end{equation}
and $\theta$ as in \eqref{theta} and
\begin{equation}\label{tildetheta}
\tilde\theta = \frac{\sqrt{9+4 \pi ^2 \left(-9+4 A \left(-3+4 A \pi ^2\right)\right)}}{16 A \pi ^2}.
\end{equation}

We then have a general solution of the form
\begin{equation}\label{Solution2}
\begin{array}{rl}
v(x) = & C_1 \,  h_1(x)  \,\, 
{}_2 F_1 (\theta+\tilde\theta, 1+\theta+\tilde\theta, 1+2\tilde\theta, \gamma(x)) \\[3mm]
+ & C_2 \, h_1(x) h_2(x) \, 
\,\, {}_2 F_1(\theta-\tilde\theta,1+\theta -\tilde\theta,
1-2\tilde\theta, \gamma(x) ),
\end{array}
\end{equation}
with $\gamma(x)$ as in \eqref{deltax},
which is again given in terms of  Gauss hypergeometric functions.

\begin{figure}
\begin{center}
\includegraphics[scale=0.85]{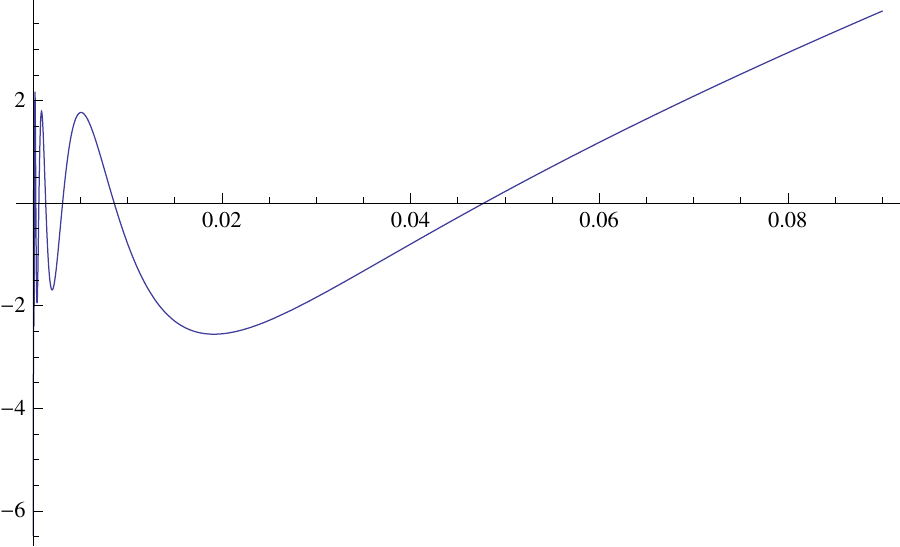}
\caption{A solution $v(x)$ with $a_\lambda = 32 \pi \rho_0 a$ and with
$A=9/(32 \pi^2)$ and $C=2$, with oscillatory behavior. 
\label{solutionoscillatoryFig}}
\end{center}
\end{figure}

One advantage with respect to the family of solutions discussed
in the previous section  is that now the value $A=\frac{9}{32\pi^2}$
of \eqref{aAy} is compatible with a positive anomalous dimension
$a_\lambda$. In this case again one can construct examples of
solutions $u(t)=v(e^{2 at})$ for which the resulting function
$\lambda(t)$ as in \eqref{lambdaSol} is continuous and positive
for all $t\geq 0$.

\medskip
\subsection{Implementing a Shaposhnikov--Wetterich-type Higgs mass estimate}

The Higgs mass estimate of 126 GeV derived 
by  Shaposhnikov Wetterich in \cite{ShaWett}, uses the RGE with anomalous
dimensions \eqref{yRGE} and \eqref{lambdaRGE}, with values of the
parameters $A=9/(32 \pi^2)$, $a= |a_y|/(16 \pi \rho_0)$ with
$\rho_0\sim 0.024$ and $|a_y|\sim 0.155$,  $C = \frac{a}{B^2}-A$, with $B=y(0)$, 
where the value $y(0)$ is constrained physically by 
compatibility with the top quark mass.
In \cite{ShaWett} they use a value of $a_\lambda \sim 3.1$ 
for the anomalous dimension, though they note that their
qualitative argument does not depend on the precise values
of the anomalous dimensions. Here we work with the explicit
solutions in terms of hypergeometric functions, which are
instead very sensitive to the explicit values of $a_\lambda$
and $a_y$. We show here that a slightly larger value of $a_\lambda\sim 5.08$
is compatible with the value of $|a_y|\sim 0.155$ used 
in  \cite{ShaWett} and with the constraint on $y(0)$ imposed
by compatibility with the top quark mass. In particular, 
for $m_t \sim 171.3$ GeV, and the Higgs vacuum $v_H \sim 246$ GeV,
which means a value $y(0)\sim 0.98$, and keeping the
same value of $A=9/(32 \pi^2)$, we obtain $C\sim 0.09$. 

\smallskip

\begin{figure}
\begin{center}
\includegraphics[scale=0.85]{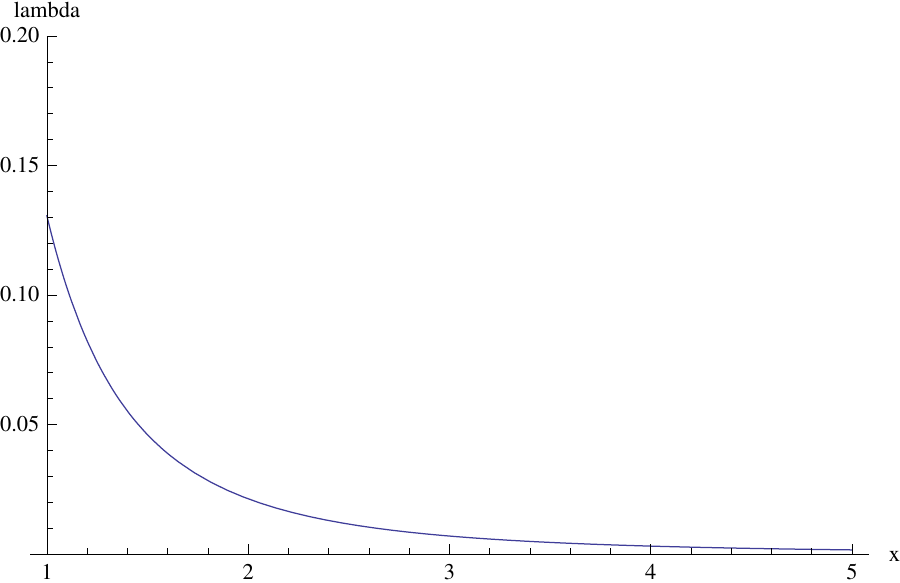}
\caption{A solution $\lambda(x)$ with $a_\lambda = 5.08$ and with
$A=9/(32 \pi^2)$, compatible with a top quark mass of $m_t \sim 171.3$ GeV
and giving a Higgs mass $m_H\sim 125.4$.
\label{lambdaxHiggsFig}}
\end{center}
\end{figure}

We have a solution of the equation  \eqref{lambdaRGE}
of the form \eqref{lambdaSol}, with $u(t)=v(e^{2a t})$ and $v(x)$
a solution of \eqref{SingODEx} of the form
\begin{equation}\label{vxBC}
v(x)= \Re(x^\sigma \left(\frac{9}{32 C \pi ^2}+x\right)^\omega
\left(9+32 C \pi ^2 x\right)^{2/3}\, {}_2 F_1(r,s,q, -\frac{32}{9} C \pi ^2 x)),
\end{equation}
for
$$ \sigma = \frac{1}{6} (-1-3 \Theta)+\frac{1}{6} \left(3+\sqrt{-12 \Theta+9 \Theta^2+\frac{1}{36} \left(144-2304 \pi ^2\right)}\right) $$
$$ \begin{array}{rl}
\omega=& \displaystyle{\frac{1}{2} (1+\frac{1}{6} (-3+3 \Theta+\sqrt{4-12 \Theta+9 \Theta^2-64 \pi ^2}-i \sqrt{-49+64 \pi ^2})}  \\[3mm] +&
\displaystyle{\frac{1}{3} (-3-\sqrt{-12 \Theta+9 \Theta^2+\frac{1}{36} (144-2304 \pi ^2)})} \\[3mm]
+ &  \displaystyle{\frac{1}{12}(18-6 \Theta-i \sqrt{-196+256 \pi ^2}+2 \sqrt{-12 \Theta+9 \Theta^2+\frac{1}{36} (144-2304 \pi ^2)})) } 
\end{array}
$$
$$ r=\frac{1}{6} \left(-3+3 \Theta+\sqrt{4-12 \Theta+9 \Theta^2-64 \pi ^2}-i \sqrt{-49+64 \pi ^2}\right) $$
$$ s=\frac{1}{12} \left(18-6 \Theta-i \sqrt{-196+256 \pi ^2}+2 \sqrt{-12 \Theta+9 \Theta^2+\frac{1}{36} \left(144-2304 \pi ^2\right)}\right) $$
$$ q=\frac{1}{3} \left(3+\sqrt{-12 \Theta+9 \Theta^2+\frac{1}{36} \left(144-2304 \pi ^2\right)}\right) .$$
Here the parameter $A$ has been assigned the value $A=9/(32 \pi^2)$, while, 
as above, the parameter $C$ is given by $C = \frac{a}{B^2}-A$, with $B=y(0)$
and the parameter $\Theta$ is given by imposing a relation of the form
$a_\lambda = 32 \pi \rho_0 a \Theta$. 
With the parameters assigned as above, we find that $\lambda(t)$
is continuous and positive for all $t\geq 0$ and such that
$\lambda(0) =0.130$, compatibly with a Higgs mass estimate
of $m_H= \sqrt{2 \lambda} v_H \sim 125.4$, for a slightly larger anomalous
dimension $a_\lambda \sim 5.08$. The behavior of this solution for
$x\geq 1$ (that is, for $t\geq 0$) is illustrated in Figure \ref{lambdaxHiggsFig},
where we plotted the function
$$ \lambda(x) = - \frac{4 \pi^2 a}{3} \, \frac{v^\prime(x)}{v(x)}. $$

\medskip
\subsection{Infrared behavior}

The solutions obtained in this way acquire 
non-physical negative values, and possibly divergences, 
at the infrared end of the spectrum.
This phenomenon was noticed also in \cite{ShaWett} and related
to radiatively induced spontaneous symmetry breaking of 
Coleman--Weinberg type \cite{CoWei}. In fact, the problem
with the behavior of solutions for large negative $t<0$ can
simply be attributed to the failure, in this range, of the approximations
\eqref{yRGE} and \eqref{lambdaRGE}, where the terms with the coupling constants
should be reintroduced as in \eqref{betalambda} and \eqref{betatop}, with
explicit solutions for the coupling constants of the form \eqref{yprimeyy3sol}.

\section{Geometric conditions at unification}\label{UnifSec}

In the models of particle physics based on noncommutative geometry
and the spectral action functional, there are several constraints on the
parameters at unification energy, imposed by the geometry of the model.
We recall here the constraints derived within the model of \cite{CCM}
and we discuss how these are affected by the presence of the
gravitational correction terms discussed above in the renormalization
group flow.

\medskip
\subsection{Constraints at unification in the CCM model}

As discussed in \cite{CCM} and in \S 1 of \cite{CoMa}, there are constraints on the
values of the parameters listed above at unification energy, imposed by the geometry
of the model. We recall them here and we refer the reader to the references above for
a detailed account of how one obtains these relations. We then show that, in the
presence of anomalous dimensions the geometric conditions at unification can
be expressed in terms of the presence of a Gaussian matter fixed point  
$F_{UV}=(\cA_F,\cH_F,0)$ for the flow on the moduli space $\tilde\cM_F$ of
the finite geometries, and this changes the  interpretation of the constraints at
unification energy with respect to  \cite{CCM} and \cite{CoMa}.
In particular, the Majorana mass terms become compatible with the
dark matter models of Shaposhnikov and Tkachev \cite{Shap1}, \cite{Shapo}, 
\cite{ShapoTka}, but one loses the geometric see-saw mechanism of \cite{CCM}.

\subsubsection{Coupling constants} The normalization of the Yang--Mills terms in
the asymptotic expansion of the spectral action (\cite{CCM}) leads to a unification
of the coupling constants with the common value $g$ at unification energy satisfying
\begin{equation}\label{coupleU}
\frac{g^2 f_0}{2 \pi^2} = \frac{1}{4} \ \ \ \Rightarrow \ \  \frac{\pi^2}{f_0} = 2 g^2 . 
\end{equation}
This fixes the value of the parameter $f_0$ of the model in terms of
the value at unification of the coupling constants $g^2=g_3^2|_{\Lambda_{{\rm unif}}}=
g_2^2|_{\Lambda_{{\rm unif}}}= (5/3) g_1^2|_{\Lambda_{{\rm unif}}}$.

\subsubsection{Higgs vacuum} The parameter $\fa$ is related to the Higgs
vacuum $\upsilon_H$, through the relation
\begin{equation}\label{MW}
\frac{\sqrt{\fa f_0}}{\pi} = \upsilon_H,
\end{equation}
which follows from a direct comparison of the terms involving the Yukawa coupling
in the asymptotic expansion of the spectral action and in the Lagrangian of
the Standard Model (see (4.13) and (4.15) of \cite{CCM}).

\subsubsection{Mass relation} The two relations above are used in \cite{CCM}
to derive an additional quadratic mass relation at unification, which follows again
from  (4.13) and (4.15) and the expression $\upsilon_H= 2 M_W/g$ for the Higgs
vacuum, where $M_W$ the $W$-boson mass. One obtains then, using \eqref{coupleU} and
\eqref{MW},
\begin{equation}\label{massrelU}
\sum_{{\rm generations}} (m_{\nu}^2 + m_e^2 + 3 m_u^2 +3 m_d^2)|_{\Lambda=\Lambda_{{\rm unif}}} = 8 M_W^2|_{\Lambda=\Lambda_{{\rm unif}}}.
\end{equation}

\subsubsection{Higgs self-coupling} The value of the Higgs self-coupling parameter
$\lambda$ at unification energy is predicted to satisfy the relation
\begin{equation}\label{lambdaU}
\lambda(\Lambda_{{\rm unif}})= \frac{\pi^2 \, 
\fb(\Lambda_{{\rm unif}})}{2 f_0 \, \fa(\Lambda_{{\rm unif}})^2}.
\end{equation}

\subsubsection{See-saw mechanism} The see-saw mechanism for the neutrino sector,
implemented geometrically by the component of the Dirac operator on $F$ acting on
the right-handed neutrino sector of the fermion representation, imposes a condition
at unification on the parameters $\fc$, $f_2$ and $f_0$
\begin{equation}\label{cf2U}
\frac{2f_2\, \Lambda_{{\rm unif}}^2}{f_0} \leq \fc(\Lambda_{{\rm unif}}) \leq 
\frac{6 f_2 \, \Lambda_{{\rm unif}}^2}{f_0}.
\end{equation}

\subsubsection{Gravitational constant} At unification energy the (effective)
gravitational constant is related to the parameters $\fc$, $f_2$ and $f_0$
of the model through
\begin{equation}\label{GeffU}
G_{{\rm eff}}(\Lambda_{{\rm unif}})= \frac{3\pi}{192\, f_2 \Lambda_{{\rm unif}}^2 - 
2 f_0 \fc(\Lambda_{{\rm unif}})}.
\end{equation}

\subsubsection{Cosmological constant} At unification energy the (effective)
gravitational constant is related to the parameters 
of the model through
\begin{equation}\label{gammaeffU}
\gamma_{{\rm eff}}(\Lambda_{{\rm unif}})= \frac{1}{4\pi^2} (192\, f_4\, \Lambda_{{\rm unif}}^4
- 4 f_2 \Lambda_{{\rm unif}}^2 \, \fc(\Lambda_{{\rm unif}}) + f_0 \, \fd(\Lambda_{{\rm unif}})).
\end{equation}
(We use here the notation $\gamma$ for the cosmological constant, as the more
commonly used $\Lambda$ is already taken by the standard notation in NCG 
for the energy scale in the spectral action.)

\subsubsection{Conformal gravity terms}

The conformal gravity terms given by the Weyl curvature term 
$\int C_{\mu\nu\rho\sigma} C^{\mu\nu\rho\sigma}$ and
the conformal coupling of the Higgs field to gravity, appear, respectively,
with coefficients $\alpha_0=-3 f_0/(10\pi^2)$ and $\xi_0= 1/12$. The latter
is the standard coefficient for the conformal coupling of the Higgs to gravity.
These coefficients are computed after a normalization of the kinetic term
of the Higgs.

\medskip
\subsection{The case with anomalous dimensions}

When we consider renormalization group equations with
gravitational corrections given by anomalous dimensions
as in \S \ref{RGEgravSec}, these have a Gaussian matter fixed point
at high energies, where the matter fields becomes uncoupled 
and asymptotically free. This means that the coupling constants
and the Yukawa parameters tend rapidly to zero at high
energies, as we have seen in the behavior of the solutions to
the approximate RGE equations discussed in \S \ref{RGEgravSec}.
This leads to a corresponding reinterpretation of the geometric
boundary conditions at unification.

In fact, if we interpret the RGE flow as a flow of the finite
geometry $F$, the Gaussian matter fixed point corresponds
to a fixed point in the moduli space $\tilde \cM_F$ as described
in \S \ref{NCGrunSec}. This fixed point has trivial coupling
constants, and in the fiber $\cM_F$ over this point of the base
$\cB_F$ it also has trivial Yukawa couplings, so that $F_{UV}$
corresponds to the finite geometry $F_{UV}=(\cA_F,\cH_F,0)$
with trivial Dirac operator. Notice that $D=0$ is also a fixed
point of all the inner fluctuations of the geometry. Thus, the
geometric constraints at unification now become the
condition that the finite geometry is close to the 
fixed point $F_{UV}$.

\smallskip

In terms of physical consequences of these boundary conditions,
one of the primary effects is on the Majorana mass terms. 
In the setting of \cite{CCM} the top eigenvalue of the Majorana
mass terms matrix $M$ is assumed to be near unification scale,
so that the three see-saw scales fall in between the unification
and the electroweak scale (see a detailed discussion of this
point in \cite{MaPie}). This is convenient for a geometric
interpretation of the see-saw mechanism for neutrino masses
(see \cite{CCM}). In the setting with anomalous dimensions
and Gaussian matter fixed point the behavior of the Majorana
mass terms becomes instead consistent with the dark matter
models of Shaposhnikov and Tkachev \cite{Shap1}, \cite{Shapo}, \cite{ShapoTka}, where
the Majorana masses are at or below the electroweak scale,
thus making them plausible dark matter candidates, in agreement
with the exclusion estimates of Kusenko \cite{Kus}.

\smallskip

Comparing with the boundary conditions of \cite{CCM}
recalled above, we see that in the case with 
Gaussian matter fixed point  $F_{UV}=(\cA_F,\cH_F,0)$,
the relation \eqref{lambdaU} can be used to fix $f_0$, 
while \eqref{MW} fixes the shape of the quartic potential
near $F_{UV}$. The relation \eqref{cf2U} still fixes $f_2$,
but now no longer implies a see-saw mechanism for
neutrino masses, for the reason mentioned above that
the Majorana mass scales can fall below the electroweak
scale as in the Shaposhnikov--Tkachev models of 
dark matter \cite{Shap1}, \cite{Shapo}, \cite{ShapoTka}.

\section{Further questions}

Some interesting questions arise in connection to the RGE equations discussed in the
previous sections.
It would be desirable to have a full derivation of Wetterich's FRGE \eqref{RGESA},
directly expressed in terms of the non-perturbative form of the spectral action, and their
compatibility with the RGE flow of \cite{AKLRS} for the particle physics sector of the 
model, with gravitational corrections as in \eqref{betagravRGE}, \eqref{betagrav}.
In view of the recent interest in cosmological applications of the NCG models,
it would also be interesting to see how the gravitational corrections to the RGE flow affect
the RGE analysis of early universe models of \cite{MaPie} and the constraints on
the use of the Higgs as a possible inflation mechanism within this type of models,
see \cite{BFS}. A related question is whether one can obtain constraints
on the cosmic topology, via the analysis of \cite{MaPieTeh1}, \cite{MaPieTeh2},
and \cite{CaMaTeh}, and the effects of the gravitational terms on the running.
Other NCG models have been proposed recently, among them
the model with supersymmetry of \cite{BroSuij} and a model with one lighter
and one heavier Higgs \cite{Ste}. It would be interesting to see what
effect the inclusion of gravitational corrections of the form \eqref{betagravRGE}, 
\eqref{betagrav} has on the RGE analysis for these models. In particular, 
the supersymmetric model of \cite{BroSuij}, first developed only for
the QCD sector, will likely extend to the electroweak sector in a way that
is not identical to the MSSM, hence the RGE equations will likely be 
also different from those of MSSM. The Wetterich method may provide a
useful way to perform the RGE analysis of the spectral action for that
model. 

\bigskip

\bigskip

\subsection*{Acknowledgments}
The first author was supported by a Summer Undergraduate Research Fellowship. 
The second author is partially supported by NSF grants DMS-0901221, 
DMS-1007207, DMS-1201512, PHY-1205440.


\begin{thebibliography}{99}

\bibitem{AKLRS} S.~Antusch, J.~Kersten, M.~Lindner, M.~Ratz, M.A.~Schmidt 
{\em Running neutrino mass parameters in see-saw scenarios}, JHEP 03 (2005) 024, 
hep-ph/0501272v3.

\bibitem{ACKPRW} H.~Arason, D.J.~Castano, B.~Kesthlyi, E.J.~Piard, P.~Ramond, 
B.D.~Wright, {\em Renormalization-group study of the standard model and its 
extensions: the standard model}, Phys. Rev. D, 46 (1992) N.9, 3945--3965.

\bibitem{Avra} I.G.~Avramidi, {\em Covariant methods for the calculation of the effective action in quantum field theory and investigation of higher- derivative quantum gravity}, 
PhD Thesis, Moscow University, 1986, hep-th/9510140.

\bibitem{BroSuij} T.~van den Broek, W.D.~van Suijlekom, {\em Supersymmetric QCD and noncommutative geometry}, Comm. Math. Phys. 303 (2011), no. 1, 149--173.

\bibitem{BFS} M.~Buck, M.~Fairbairn, M.~Sakellariadou, {\em 
Inflation in models with conformally coupled scalar fields: An application to the noncommutative spectral action}, Phys. Rev. D 82 (2010) 043509 [14 pages]

\bibitem{Cac}  B.~\'Ca\'ci\'c, 
{\em Moduli spaces of Dirac operators for finite spectral triples}, in ``Quantum groups and noncommutative spaces", pp. 9--68, Aspects Math., E41, Vieweg Teubner, Wiesbaden, 2011.

\bibitem{CaMaTeh} B.~\'Ca\'ci\'c, M.~Marcolli, K.~Teh, {\em 
Coupling of gravity to matter, spectral action and cosmic topology},
 arXiv:1106.5473, to appear in Journal of Noncommutative Geometry.

\bibitem{CaIoSchu} L.~Carminati, B.~Iochum, T.~Sch\"ucker, 
{\em The noncommutative constraints on the standard model \`a la Connes}, J. Math. Phys. 38 (1997), no. 3, 1269--1280.

\bibitem{CaIoKaSchu} L.~Carminati, B.~Iochum, D.~Kastler, T.~Sch\"ucker, {\em Relativity, noncommutative geometry, renormalization and particle physics. Coherent states, differential and quantum geometry}, Rep. Math. Phys. 43 (1999), no. 1-2, 53--71.

\bibitem{CC2} A.~Chamseddine, A.~Connes, {\it The spectral
action principle}, Commun. Math. Phys. 186 (1997) 731--750.

\bibitem{CC3} A.~Chamseddine, A.~Connes, {\em The uncanny
precision of the spectral action}, Comm. Math. Phys. 293 (2010), no. 3, 867--897.

\bibitem{CCnew} A.~Chamseddine, A.~Connes, {\em Resilience of the Spectral Standard Model},
arXiv:1208.1030.

\bibitem{CCM} A.~Chamseddine, A.~Connes, M.~Marcolli, {\em 
Gravity and the standard model with neutrino mixing}, Adv. Theor. Math. Phys. 11 (2007), no. 6, 991--1089.

\bibitem{CoWei} S.~Coleman, E.~Weinberg, {\em Radiative corrections as the origin of spontaneous symmetry breaking}, Phys. Rev. D7 (1973) 1888--1910.

\bibitem{CoSM} A.~Connes, {\em Gravity coupled with matter and the foundation of noncommutative geometry}, Comm. Math. Phys. 182 (1996), no. 1, 155--176.

\bibitem{CoMa} A.~Connes, M.~Marcolli, {\em Noncommutative geometry, quantum
fields and motives}, Colloquium Publication, Vol.55, American Math. Soc. 2008.

\bibitem{DouPe} D.~Dou, R.~Percacci, {\em The Running Gravitational Couplings}, 
Class. Quant. Grav. 15 (1998) 3449--3468.

\bibitem{MaKol} D.~Kolodrubetz, M.~Marcolli, 
{\em Boundary conditions of the RGE flow in the noncommutative geometry approach to particle physics and cosmology}, Physics Letters B 693 (2010) 166--174.

\bibitem{Kus} A.~Kusenko, {\em Sterile neutrinos: the dark side of the light fermions}, 
arXiv:0906.2968.

\bibitem{LauReut} O.~Lauscher, M.~Reuter, {\em 
Ultraviolet fixed point and generalized flow equation of quantum gravity},
 Phys. Rev. D 65 (2002) 025013 [44 pages]

\bibitem{Mar} M.~Marcolli, {\em Building cosmological models via noncommutative geometry}, International Journal of Geometric Methods in Modern Physics, Vol.8, N.5 (2011) 1131--1168.

\bibitem{MaPie} M.~Marcolli, E.~Pierpaoli, {\em Early universe models from noncommutative geometry}, Adv. Theor. Math. Phys. 14 (2010), no. 5, 1373--1432.

\bibitem{MaPieTeh1} M.~Marcolli, E.~Pierpaoli, K.~Teh,
{\em The spectral action and cosmic topology}, Comm. Math. Phys. 304 (2011), no. 1, 125--174.

\bibitem{MaPieTeh2} M.~Marcolli, E.~Pierpaoli, K.~Teh,
{\em The coupling of topology and inflation in noncommutative cosmology}, 
Comm. Math. Phys. 309 (2012), no. 2, 341--369.

\bibitem{NaPe} G.~Narain, R.~Percacci, {\em Renormalization group flow in
scalar-tensor theories, I}, arXiv:0911.0386v2.

\bibitem{NeOcSa} W.~Nelson, J.~Ochoa, M.~Sakellariadou,
{\em Constraining the noncommutative spectral action via astrophysical observations},
Phys. Rev. Lett. 105 (2010) 101602  [5 pages]

\bibitem{NeOcSa2} W.~Nelson, J.~Ochoa, M.~Sakellariadou,
{\em Gravitational waves in the spectral action of noncommutative geometry},
Phys. Rev. D 82 (2010) 085021 [15 pages]

\bibitem{Oku} Y.~Okumura, {\em Renormalization group analysis of the Higgs boson mass in a noncommutative differential geometry}, Prog. Theor. Phys. 98 (1997) 1333--1344.

\bibitem{PePe} R.~Percacci, D.~Perini, {\em Asymptotic safety of gravity coupled
to matter}, hep-th/0304222.

\bibitem{Reut} M.~Reuter, {\em Nonperturbative evolution equation for quantum gravity},
Phys Rev D 57 (1998) N.2, 971--985.

\bibitem{Saad} N.~Saad, R.L.~Hall, H.~Ciftci, {\em Criterion for polynomial solutions
to a class of linear differential equations of second order}, J. Phys. A: Math. Gen.
39 (2006) 13445--13454.

\bibitem{Shap1} M.~Shaposhnikov, {\em The $\nu$MSM, dark matter and
baryon asymmetry of the Universe}, Journal of Physics: Conference
Series 39 (2006) 9--11.  

\bibitem{Shapo} M.~Shaposhnikov, {\em The $\nu$MSM, leptonic asymmetries, and 
properties of singlet fermions}, JHEP (2008) 0808:008, 56pp.

\bibitem{ShapoTka} M.~Shaposhnikov, I.~Tkachev, {\em The $\nu$MSM, inflation, and dark matter},
Phys.Lett. B Vol.639 (2006) 414--417.

\bibitem{ShaWett} M.~Shaposhnikov, C.~Wetterich, {\em Asymptotic safety of 
gravity and the Higgs boson mass}, arXiv:0912.0208v2.

\bibitem{Souma} W.~Souma, {\em Non-trivial ultraviolet fixed point in quantum gravity},
Prog. Theor. Phys. 102 (1999) 181--195. 

\bibitem{Ste} C.A.~Stephan, {\em New scalar fields in noncommutative geometry},
Phys. Rev. D 79 (2009) 065013 

\bibitem{Teh} K.~Teh, {\em Nonperturbative Spectral Action of Round Coset Spaces of $SU(2)$},
 arXiv:1010.1827, to appear in Journal of Noncommutative Geometry.

\bibitem{Wein} S.~Weinberg, {\em Ultraviolet divergences in quantum theories of
gravitation}, in ``General Relativity: an Einstein centenary survey" (S.W.~Hawking, 
W.~Israel, Eds.) Cambridge Univ. Press, 1979, pp.790--831.

\bibitem{Wett} C.~Wetterich, {\em Exact evolution equation for the effective potential},
Phys. Lett. B 301 (1993) 90--94.

\bibitem{ZaZaVaPe} O.~Zanusso, L.~Zambelli, G.P.~Vacca, R.~Percacci,
{\em Gravitational corrections to Yukawa systems}, arXiv:0904.0938v2.

\end{thebibliography}
\end{document}